\documentclass[aps,prb,twocolumn,showpacs]{revtex4}
\usepackage{bm}
\usepackage{graphicx}
\usepackage{amsmath}
\usepackage{eufrak}
\usepackage{color}
\newcommand{\nix}[1]{}

\begin{document}

\title{Orbital photogalvanic effects in quantum-confined structures}
\author{J.~Karch,$^1$ S.A.~Tarasenko,$^2$ P.~Olbrich,$^1$ T.~Sch\"onberger,$^1$ 
C.~Reitmaier,$^1$ D.~Plohmann,$^1$ Z.D.~Kvon,$^3$ and S.D.~Ganichev$^{1}$}
\affiliation{$^1$Terahertz Center, University of Regensburg,
93040 Regensburg, Germany}
\affiliation{$^2$A.F.\,Ioffe Physical-Technical Institute, Russian Academy of
Sciences, 194021 St.\,Petersburg, Russia}
\affiliation{$^3$ Institute of Semiconductor Physics, Russian Academy
of Sciences, 630090 Novosibirsk, Russia}

\begin{abstract}
We report on the circular and linear photogalvanic effects caused by free-carrier 
absorption of terahertz radiation in electron channels on (001)-oriented and miscut 
silicon surfaces. The photocurrent behavior upon variation of the radiation polarization 
state, wavelength, gate voltage and temperature is studied. We present the microscopical and 
phenomenological theory of the photogalvanic effects, which describes well the experimental 
results. In particular, it is demonstrated that the circular (photon-helicity sensitive) photocurrent in 
silicon-based structures is of pure orbital nature originating from the quantum interference of
different pathways contributing to the absorption of monochromatic radiation.
\end{abstract}

\maketitle

\section{Introduction}

The photogalvanic effects, linear (LPGE) and circular (CPGE), representing the generation 
of a directed electric current due to the asymmetry of photoexcitation or relaxation processes attract 
growing attention. They can occur in media of sufficiently low spatial symmetry only and are known to be 
an efficient tool to study nonequilibrium processes in semiconductor structures yielding
information about their symmetry, details of the band spin splitting, momentum, energy and spin relaxation
times etc. (see e.g. [\onlinecite{SturmanFridkin_book,Ivchenko_book,GanichevPrettl,IvchenkoGanichev_book}]).
Moreover, the photogalvanic effects are applied for the detection of the
polarization state of infrared and terahertz (THz) laser radiation.~\cite{JAP2009}  
The low spatial symmetry required for the photogalvanic effects is naturally realized in 
non-centrosymmetric semiconductors and low-dimensional structures based on them.
However, it has been shown that both the LPGE and the CPGE
can occur in two-dimensional structures made of centrosymmetric crystals, e.g. silicon, 
due to the structure inversion asymmetry.~\cite{Magarill89,IvchenkoPikus_book,Tarasenko07}
While the LPGE in silicon metal-oxide-semiconductor field-effect-transistors (Si-MOSFETs) was 
observed almost twenty years ago,~\cite{Gusev87} first experiments on the CPGE have been reported 
only most recently.~\cite{Olbrich09}
Besides the fundamental interest to photogalvanics, a renaissance
in the study of photoelectrical phenomena in MOSFETs is additionally stimulated by promising
applications for the emission and detection of terahertz radiation.~\cite{Knap09}   
Here, we present a comprehensive experimental and theoretical study of the THz radiation induced 
photogalvanic effects in Si-MOSFETs fabricated on (001) precisely oriented as well as on miscut surfaces.  
The fact of the existence of the circular (photon-helicity sensitive) photocurrent, 
which reverses its direction upon changing the sign of the circular polarization,
in Si-based structures is of particular interest. It demonstrates that the CPGE current  
caused by the transfer of photon angular momenta to free carriers can be generated 
not only in structures with a strong spin-orbit interaction due to spin-dependent optical excitation,~\cite{Ganichev01,GanichevPrettlreview,Bieler05}
but even in systems with a vanishingly small constant of spin-orbit coupling. We show that the CPGE in
Si-MOSFETs excited by THz radiation originates from the quantum interference of different pathways 
contributing to the radiation absorption.\,\cite{Tarasenko07,Olbrich09}

This paper is organized as follows.
In Sec.~\ref{sphenomen}, we present the phenomenological theory of the LPGE and CPGE in two-dimensional channels
on (001) precisely oriented structures as well as on miscut silicon surfaces. 
In Sec.~\ref{sexperiment}, a short overview of the used experimental technique is given.
The experimental results are summarized in Sec.~\ref{sresults} and compared with the phenomenological theory. 
In Sec.~\ref{stheory}, we present a microscopic theory of the CPGE caused by the quantum interference of optical 
transitions for both types of structures. Finally, in Sect.~\ref{sdiscussion},
we discuss the experimental data in view of the theoretical background.

\section{Phenomenological theory}\label{sphenomen}

Phenomenologically, the density of the photocurrent $\bm{j}$ emerging in an unbiased 
structure within the linear regime in the radiation intensity  is given by~\cite{IvchenkoPikus_book}
\begin{equation} \label{phen0}
j_\alpha = \sum_{\beta\gamma}
\chi_{\alpha\beta\gamma}\:\frac{\left(E_\beta E^*_\gamma +
E_\gamma  E^*_\beta\right)}{2}
+ \sum_{\beta} \mu_{\alpha\beta}\, i [\bm{E} \times \bm{E}^*]_\beta \:,
\end{equation}
where $\bm{\chi}$ is a third-rank tensor being symmetric in the last two indices,
$\bm{\mu}$ is a second-rank pseudotensor, 
$\bm{E} = |\bm{E}| \bm{e}$ is the electric field of the radiation  in the channel, 
$\bm{e}$ is the (complex) unit vector of the light polarization, 
$\alpha$, $\beta$ and $\gamma$ are the Cartesian coordinates, and the photon wave 
vector is neglected in Eq.~(\ref{phen0}). The 
tensor $\bm{\chi}$ describes the photocurrent, which 
can be induced by linearly polarized or unpolarized radiation. 
By contrast, $\bm{\mu}$ stands for the photocurrent, which is sensitive to the radiation
helicity  and  reverses its
direction upon switching the sign of the circular polarization, because
$i [\bm{E} \times \bm{E}^*] = |\bm{E}|^2 P_{circ} \hat{\bm{e}}$ with the degree of circular polarization $P_{circ}$ 
and the unit vector $\hat{\bm{e}}$ pointing in the light propagation direction.
Non-zero components of the tensors $\bm{\chi}$ and $\bm{\mu}$ can be 
obtained by analyzing the spatial symmetry of structures.

We consider the linear and circular photogalvanic effects
for the two point-group symmetries $C_{\infty v}$ and $C_s$, which are relevant to electron 
channels on precisely (001) oriented and miscut silicon surfaces, respectively. 
Two-dimensional channels on a (001) surface 
can be effectively described by the axial point-group $C_{\infty v}$,
which takes into account the structure inversion asymmetry of the channel. 
In such structures, the photocurrent can be excited only at oblique incidence of radiation,
and its components are given by the phenomenological equations
\begin{eqnarray}\label{phen001}
j_x  = L (E_x E_z^* + E_z E_x^*) +  C \, i [\bm{E} \times \bm{E}^*]_y \:, \\
j_y  = L (E_y E_z^* + E_z E_y^*) - C \, i [\bm{E} \times \bm{E}^*]_x \:, \nonumber
\end{eqnarray}
where the first and second terms on the right-hand side of Eqs.~(\ref{phen001}) stand for 
the linear and circular photogalvanic effects, respectively, 
$x$ and $y$ are the in-plane axes, and $z \parallel [001]$ is the channel normal.

The channels fabricated on miscut surfaces exhibit the lower point-group symmetry $C_s$
due to the deviation of the channel plane from $(001)$ together with 
the asymmetry of the confinement potential.~\cite{Magarill89,IvchenkoPikus_book}
In structures of the $C_s$ point group, both linear and circular photogalvanic effects are allowed
even at normal incidence of radiation.~\cite{GanichevPrettlreview}
In this particular geometry, the photocurrent components are phenomenologically described by
\begin{eqnarray}\label{phen_vicinal}
j_{x'} &=& L'_1 (E_{x'} E_{y'}^* + E_{y'} E_{x'}^*) + C' \, i [\bm{E} \times \bm{E}^*]_{z'}  \:, \\
j_{y'} &=& L'_2 + L'_3 (|E_{x'}|^2 - |E_{y'}|^2) \:. \nonumber
\end{eqnarray}
Here, we assume that the channel plane $(x'y')$ is tilted
from the plane $(001)$ around the axis
$x' \parallel [1\bar{1}0]$, and $z'$ is the channel normal. In contrast to $(001)$-oriented MOSFETs, 
the photocurrent in structures on miscut surfaces can be induced even by unpolarized radiation.

\section{Samples and experimental techniques}\label{sexperiment}

In our experiments, we study photocurrents in MOSFETs
fabricated on (001)-oriented and miscut silicon surfaces.
On the precisely (001)-oriented surface, a transistor along $y \parallel [110]$ 
with a channel length of $3$\,mm and a width of $2.8$\,mm
was prepared with a $110$\,nm thick SiO$_2$ layer, a semitransparent polycrystalline Si gate
and a doping level $N_a$ of the depletion layer of about $3 \times 10^{15}$\,cm$^{-3}$.
In this transistor, the variation of the gate voltage V$_g$ from $1$ to $20$\,V changes 
the carrier density $N_s$ from
about $1.9 \times 10^{11}$ to $3.8 \times 10^{12}$\,cm$^{-2}$ and the energy spacing 
$\varepsilon_{21}$ between the size-quantized subbands $e1$ and $e2$ from $10$ to $35$\,meV.
The peak mobilities $\mu$ at room and helium temperature are $700$ and $10^4$\,cm$^2$/Vs,
respectively.
Another set of MOSFETs was fabricated on miscut surfaces tilted by an angle of $\vartheta=9.7^\circ$
or $\vartheta=10.7^\circ$ from the $(001)$-plane around $x^\prime\parallel [1\bar{1}0]$.
Two transistors were prepared on each miscut substrate, one oriented along $x^\prime$ and the other along the inclination 
direction $\bm A \parallel y^\prime$. They have a size of $1.2\times0.4$~mm$^2$, semitransparent 
Ti gates of $10$\,nm thickness and a doping concentration $N_a$ of about $1 \times 10^{13}$\,cm$^{-3}$.
A variation in the gate voltage from 1 to $20$\,V changes the carrier density $N_s$ from 
$1.5 \times 10^{11}$ to $3.0 \times 10^{12}$\,cm$^{-2}$ and the energy spacing $\varepsilon_{21}$ from $5$ to $28$\,meV.
The peak electron mobilities $\mu$ in the channel are about 10$^3$ and $2\times
10^4$\,cm$^2$/Vs at $T = 296$ and $4.2$\,K, respectively.
The MOSFETs on miscut surfaces have a $140$\,nm thick SiO$_2$ layer.

For optical excitation we use the emission of a terahertz molecular gas laser, 
optically pumped by a transversely excited atmosphere pressure (TEA) CO$_2$ laser.\cite{GanichevPrettl} 
With NH$_3$ as active gas, 100~ns pulses of radiation 
with a peak power $P \approx$\,30~kW are obtained at the wavelengths $\lambda=$ 76, 90, 148 and 280~$\mu$m 
(corresponding to the photon energies $\hbar \omega=$ 16.3, 13.7, 8.4 and 4.4~meV, respectively).
The terahertz radiation causes in Si-MOSFETs indirect Drude-like optical transitions 
or direct resonant intersubband transitions, which can be tuned by the gate voltage.
Various polarization states of the radiation are achieved by transmitting the linearly polarized 
laser beam through $\lambda$/2 or $\lambda$/4 crystal quartz plates.
By applying $\lambda/2$ plates the angle  $\alpha$ between the plane of linear polarization and the 
$y$ or $y^\prime$ axis is varied from $\alpha = 0^\circ$
to $180^\circ$ covering all possible orientations of the electric field vector. By rotating a $\lambda$/4 plate 
we obtain elliptically and, in particular, circularly polarized light. The polarization states are 
then described by the angle $\varphi$
between the initial polarization of the laser light ($\bm{E} \parallel y$ or $y^\prime$) and the optical axis of the plate. 
The degree of circular polarization in this set-up is given by $P_{circ} = \sin 2\varphi$.
The light polarization states for some characteristic angles $\varphi$ are sketched on the top of Fig.~\ref{figure01}(a).
The miscut and (001)-oriented samples are excited at normal and oblique incidence, respectively.
In the case of oblique incidence, the angle $\theta_0$ between the light propagation direction 
and the sample normal is $\pm 30^\circ$.
The photocurrents are measured between the source and drain contacts of the unbiased transistors via
the voltage drop across a 50$\Omega$ load resistor. The experimental geometries are
illustrated in the insets of Fig.~\ref{figure01}.

\section{Experiments}\label{sresults}

\begin{figure}[t]
\includegraphics[width=0.95\linewidth]{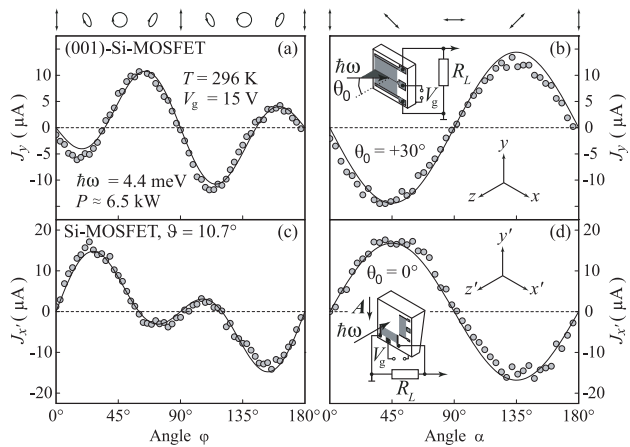}
\caption{Polarization dependences of the photocurrent measured in transistors on the
(001)-oriented surface [(a) and (b)] and on the miscut surface [(c) and (d)]. Panels (a) and (c) show the
photocurrent as a function of the angle  $\varphi$, which determines the radiation helicity.
Panels (b) and (d) show the photocurrent as a function of the azimuth angle  $\alpha$.
The data are obtained at room temperature, for a gate voltage of $15$\,V, radiation
with the photon energy $4.4$\,meV, the power $\approx 6.5$\,kW, and the diameter of the laser spot of
about $2.5$\,mm. The insets sketch the experimental set-ups for the different structures:
The (001)-MOSFET is illuminated at oblique incidence with the angle $\theta_0 = 30^{\circ}$, and the
miscut sample at normal incidence. 
On top the states of polarization of various angles $\varphi$ and $\alpha$ are illustrated.
} \label{figure01}
\end{figure}

Irradiating the MOSFET structures with THz radiation
we observed photocurrent signals with a temporal shape reproducing that of the laser pulse
of about $100$\,ns duration.
Figure~\ref{figure01} shows polarization dependences of the measured currents 
in the $y$ direction on the (001) precisely oriented surface [Figs.~\ref{figure01}(a),(b)] and in the transistor along $x^\prime$ 
on the miscut surface [Figs.~\ref{figure01}(c),(d)]. The data are obtained at room temperature, a gate voltage of $V_g = 15$\,V,
and at oblique and normal incidence of radiation with the photon energy $\hbar\omega = 4.4$\,meV 
for the former and latter structures, respectively.

First, we discuss the polarization dependences of the photocurrent in (001)-oriented structures. 
Figure~\ref{figure01}(a) shows the photocurrent $J_y(\varphi)$ measured
perpendicular to the light propagation at oblique incidence of radiation in the plane $(xz)$ for an angle $\theta_0 = 30^\circ$.
The light polarization states for several 
angles $\varphi$ are sketched on the top of Fig.~\ref{figure01}(a). Obviously the signal 
reverses its sign upon switching the radiation helicity from left- to right-handed circular polarization.
This is the characteristic fingerprint of the circular photogalvanic effect.~\cite{SturmanFridkin_book,Ivchenko_book,GanichevPrettl,IvchenkoGanichev_book}
In addition, the photocurrent has a contribution independent of the photon helicity. 
The whole dependence $J_y(\varphi)$ can be well fitted by
\begin{equation}
\label{J_phen1}
J(\varphi) = (J_L/2)\sin{4\varphi} + J_C\sin{2\varphi} \:
\end{equation}
with comparable parameters $J_L$ and $J_C$. 
Such a dependence is in full agreement
with the phenomenological Eq.~(\ref{phen001}) since the polarization dependent terms are reduced to 
$(E_y E_z^* + E_z E_y^*)  = - (E_0^2 \, t_p t_s/2) \sin \theta \sin 4\varphi$ and 
$i [\bm{E} \times \bm{E}^*]_x = E_0^2 \, t_p t_s \sin \theta \sin 2\varphi$ in this experimental 
geometry.  Here, $E_0$ is the electric field amplitude of
the incident light, $t_p$ and $t_s$ are the amplitude transmission coefficients for
$s$- and $p$-polarized radiation, $\theta$ is the angle of refraction related to the incidence angle 
by $\sin \theta = \sin \theta_0 /n_{\omega}$ with the refraction index of the medium $n_{\omega}$.
The first and second terms on the right-hand side of Eq.~(\ref{J_phen1}) 
correspond to the linear and circular photogalvanic effects, respectively, see Sec.~\ref{sphenomen}. 
Varying the angle of incidence $\theta_0$ from $-30^\circ$ to $+30^\circ$ we observed that 
both contributions to the photocurrent reverse their signs. 
Moreover, by changing the plane of incidence from $(xz)$ to $(yz)$, 
we obtained that the photon-helicity current in the $y$ direction vanishes and only the LPGE contribution remains. 
All these features correspond to the phenomenological Eq.~(\ref{phen001}).

\begin{figure}[t]
\includegraphics[width=0.7\linewidth]{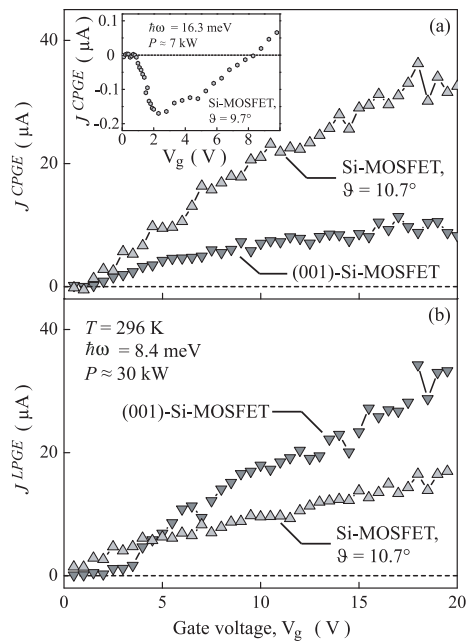}
\caption{Gate voltage dependences of (a) the circular and (b) the linear photogalvanic currents
measured in the transistors on the (001)-oriented for $\theta_0 = 30^{\circ}$ and the miscut surface with $\vartheta=10.7^{\circ}$ for normal incidence.
The data are obtained at room temperature for
the photon energy $8.4$\,meV, the power $\approx 30$\,kW, and the diameter of the laser spot of
about $2$\,mm. The inset shows the behavior of the circular photocurrent for the photon energy 
$16.3$\,meV in the miscut MOSFET with the inclination angle $9.7^{\circ}$.
} \label{figure03}
\end{figure}

Equation~(\ref{phen001}) demonstrates that the photocurrent can also be induced by linearly polarized radiation. In this case, 
the signal should solely be caused by the LPGE, because the CPGE vanishes for linear polarization. The dependence of $J_y$ on the 
azimuth angle $\alpha$ is given in Fig.~\ref{figure01}(b) and can be well fitted by
\begin{equation}
\label{J_phen2}
J(\alpha) = J_L \sin{2\alpha}
\end{equation}
with the same parameter $J_L$ used in Eq~(\ref{J_phen1}). This behavior is also expected 
from Eq.~(\ref{phen001}), because $(E_y E_z^* + E_z E_y^*) = - E_0^2 \, t_p t_s \sin \theta \sin 2\alpha$.

Now we turn to the transistors made on miscut surfaces. In contrast to (001)-oriented structures,
here we observed a photocurrent even at normal incidence of radiation [Figs.~\ref{figure01}(c),(d)]. This difference
follows from the symmetry arguments discussed above, see Eqs.~(\ref{phen001}) and~(\ref{phen_vicinal}).
The photon-helicity dependent current is detected only in the transistor along $x^\prime \parallel [1\bar{1}0]$, 
therefore, we focus below on this particular geometry. The polarization dependences $J_{x^\prime}(\varphi)$ and $J_{x^\prime}(\alpha)$
can also be fitted by Eqs.~(\ref{J_phen1}) and~(\ref{J_phen2}), respectively, with another set of fitting parameters. 
They are in accordance with the phenomenological Eq.~(\ref{phen_vicinal}) 
since at normal incidence   $(E_{x'} E_{y'}^* + E_{y'} E_{x'}^*) = -(1/2)(E_0 \,t_s)^2 \sin 4\varphi$, $i[\bm{E}\times\bm{E}^*]_{z'} = - (E_0 \,t_s)^2 \sin 2\varphi$, and $(E_{x'} E_{y'}^* + E_{y'} E_{x'}^*) = -(E_0 \,t_s)^2 \sin 2\alpha$.
However, we emphasize that, in contrast to (001)-oriented 
structures, the photocurrent in miscut samples is determined by the crystallographic axes rather than by the light propagation direction.

\begin{figure}[t]
\includegraphics[width=0.95\linewidth]{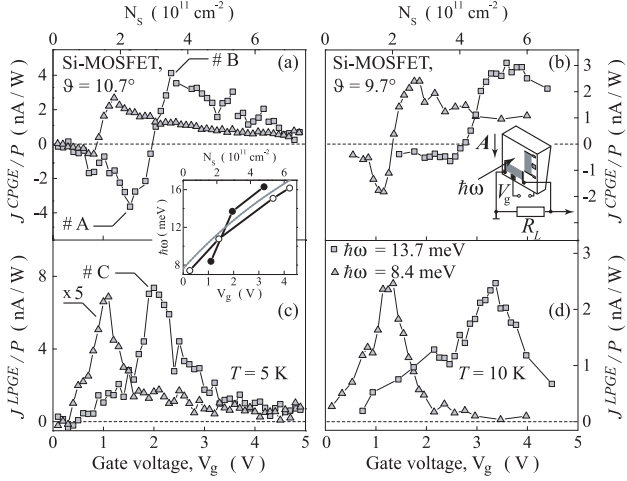}
\caption{Gate voltage dependences of the circular photogalvanic current [Figs. (a) and (b)] and linear photogalvanic current
[Figs. (c) and (d)] measured in transistors on miscut surfaces with different inclination angles normalized to the laser power.
The data are obtained at low temperatures with the photon energies $8.4$\,meV and $13.7$\,meV. 
The letters $\#$A, $\#$B and $\#$C denote the gate voltages, at which the temperature dependences of Fig.\,\protect\ref{figure05}
are measured.  Full circles in the inset show the relation between the photon energy and gate voltage of the inversion point of the CPGE.
The open circles and solid curve in the inset show the relation between the photon energy and gate voltage corresponding to the intersubband resonance
in (001)-oriented MOSFETs obtained by absorption measurements\,\protect\cite{Knesch1} and calculated numerically,\,\protect\cite{Ando1} respectively.
} \label{figure04}
\end{figure}

The contributions of the circular and linear photogalvanic effects persist for all
applied gate voltages from $1$ to $20$\,V, photon energies and temperatures used in our experiments on both types of MOSFETs.
First we discuss the gate voltage dependences of the CPGE [Fig.~\ref{figure03}(a)] and LPGE [Fig.~\ref{figure03}(b)] contributions, i.e. 
$J_C = [J(\varphi=45^{\circ}) - J(\varphi=135^{\circ})]/2$ and $J_L = [J(\alpha=45^{\circ}) - J(\alpha=135^{\circ})]/2$, respectively,
obtained at room temperature. For the photon energy $\hbar\omega = 8.4$\,meV, as well as the most other studied wavelengths, an increase in
the gate voltage results in a rise of the photocurrent magnitude. Such a behavior can be attributed to the increase 
in the electron density in the inversion channel and, therefore, the enhancement of the Drude absorption. However, while the LPGE contribution
always increases with $V_g$, the CPGE at some photon energies exhibits a sign inversion with raising $V_g$, as demonstrated for the photon energy
$\hbar\omega = 16.3$\,meV in the inset in Fig.~\ref{figure03}(a). We note that we can not attribute this gate voltage to any characteristic 
energy in the band structure of silicon-based quantum-confined channels.

\begin{figure}[t]
\includegraphics[width=0.7\linewidth]{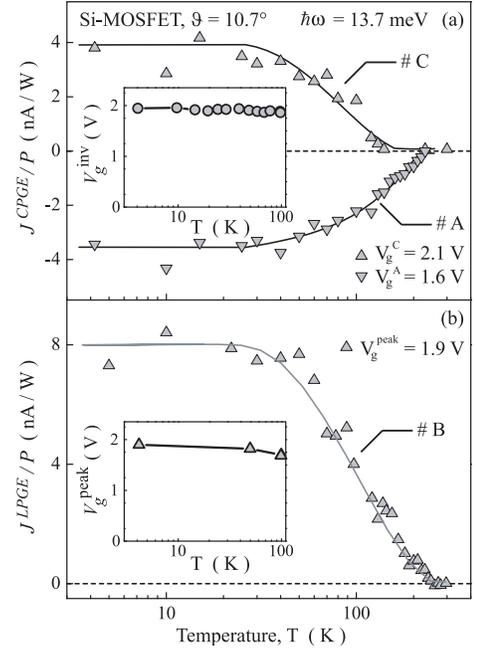}
\caption{Temperature dependences of (a) the circular and (b) the linear photogalvanic currents measured in the transistor
on the miscut surface with $\vartheta=10.7^{\circ}$ for $\hbar \omega = 13.7$\,meV
at characteristic gate voltages indicated by the letters $\#$A, $\#$B and $\#$C in Fig.\,\protect\ref{figure04}.
The insets demonstrate that the gate voltage corresponding to the CPGE sign inversion $V_g^{\rm{inv}}$ as well as the gate voltage
corresponding to the LPGE peak $V_g^{\rm{peak}}$ are almost independent of the temperature.
}\label{figure05}
\end{figure}

The most spectacular discrepancy in the CPGE and LPGE behavior emerges at helium temperature.
While the polarization dependences remain the same, the gate voltage dependences get consistently
different. In contrast to the smooth behavior observed at room temperature, at helium temperature 
the LPGE current exhibits a resonance-like response and the CPGE photocurrent shows always a sign inversion.
This is shown in Fig.~\ref{figure04} for two miscut Si-MOSFETs with different angles of inclination
$\vartheta=10.7^{\circ}$ and $9.7^{\circ}$.
The peak position of the LPGE current, which coincides with the inversion point of the CPGE, 
depends on the photon energy $\hbar\omega$ and corresponds to $\hbar\omega \approx \varepsilon_{21}$, 
where $\varepsilon_{21}$ is the energy separation of the first two electron subbands. The latter is in accordance with band structure
calculations~\cite{Ando1} and experiments on far infrared absorption~\cite{Knesch1} (see inset to Fig.~\ref{figure04}) 
as well as with our photoconductivity measurements (not shown). The resonance condition is obtained by tuning the energy separation 
between the first two electron subbands $\varepsilon_{21}$ to the
photon energy $\hbar\omega$ by changing the gate voltage: The increase in the photon energy shifts the intersubband resonance
to larger gate voltages.\cite{Stern1} The discrepancy in the resonance positions for a fixed photon energy observed in 
the two miscut samples is attributed to the difference of their inclination angles.

By increasing the temperature from liquid helium to room temperature the resonant peak of the LPGE and as well as the sign inversion of the CPGE at
$\hbar\omega \approx \varepsilon_{21}$ vanish. This is shown in Fig.~\ref{figure05} for three characteristic gate voltages:
All three signals remain almost constant below about $35$\,K and decrease for higher temperatures. The insets in Figs.~\ref{figure05}(a),(b) 
demonstrate that the inversion point of the CPGE $V_g^{\rm{inv}}$ and the peak position of the LPGE $V_g^{\rm{peak}}$ do not shift with
temperature as far as they are detectable in agreement with the behavior of $\varepsilon_{21}(T)$ well known for MOSFETs.\cite{Stern1} 
At $T > 120$\,K, the gate voltage dependences of the current contributions become smooth
similar to those observed at room temperature.

\section{Microscopic theory of the CPGE} \label{stheory} 

The microscopic theory of the photogalvanic effects caused by intersubband optical transitions
in silicon-based low-dimensional structures was developed in Ref.~[\onlinecite{Magarill89}]. This theory describes well
the resonance behavior of the linear photogalvanic effect measured at low temperatures. 
The observation of the photogalvanic effects apart the intersubband resonance demonstrates that the free
carrier absorption also gives rise to a polarization dependent photocurrent. Below we present the 
microscopic theory of the CPGE in both (001)-oriented and miscut Si-MOSFETs.
As addressed above, the generation of helicity dependent photocurrents in silicon-based structures 
is of particular interest, because spin-related mechanisms of the CPGE are ineffective due to
the vanishingly small constant of spin-orbit coupling in silicon.
Therefore, less studied pure orbital mechanisms
determine the current formation. We consider theoretically this process following 
Ref.~[\onlinecite{Tarasenko07}] and calculate the photocurrent taking into 
account the peculiarity of the silicon band structure. 

\begin{figure}[t]
\includegraphics[width=0.95\linewidth]{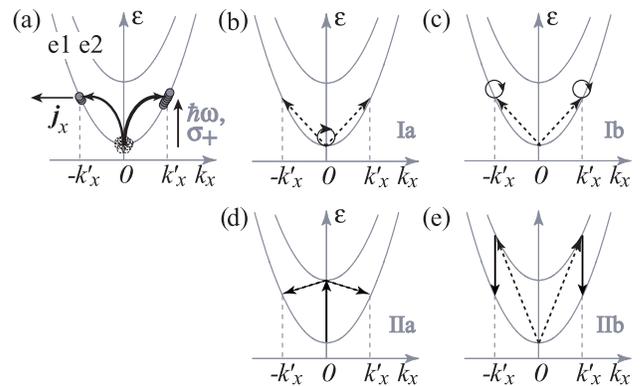}
\caption{Microscopic model of the CPGE. 
Panel (a): Indirect optical transitions due to free-carrier absorption of circularly polarized radiation
are shown by bend arrows of various thickness indicating the
difference in transition rates, which is caused by the quantum interference of various absorption pathways. 
Full circles sketch the resulting imbalance of the
carrier distribution in $\bm{k}$-space yielding an electric current $j_x$. 
Panel (b) –- (e): various pathways of the radiation absorption involving intermediate states in the subbands $e1$ and $e2$.
Here, solid arrows indicate electron-photon interaction
and the dashed arrows describe scattering events.
}\label{Figure06}
\end{figure}

We remind that the electron subbands in Si structures are formed by six valleys 
$X$, $X'$, $Y$, $Y'$, $Z$ and $Z'$ located at the $\Delta$
points of the Brillouin zone of the bulk crystal.
The electron dispersion in each valley is strongly anisotropic and described by two different 
effective masses: longitudinal $m_{\parallel}$ and transverse $m_{\perp}$ with respect to the 
principle axis of the valley. Due to quantum confinement, only two valleys,
$Z$ and $Z'$, are occupied at low temperatures and low electron densities in structures grown 
along $z \parallel [001]$, while the other four valleys are considerably
higher in energy. The low-energy valleys are almost equivalent and
can be treated independently, because the valley splitting caused
by the asymmetry of the confining potential is negligible in
comparison to the kinetic energy of electrons and because inter-valley
scattering is much weaker than intra-valley processes.

\subsection{The CPGE in (001)-oriented structures}

The model illustrating the generation of the photocurrent due to free-carrier absorption 
of circularly polarized radiation is sketched in Fig.\,\ref{Figure06}. 
Figure\,\ref{Figure06}(a) shows indirect optical transitions
within the subband $e1$ (we assume that only the ground subband 
is populated in the equilibrium). For fulfilling the conditions of energy 
and momentum conservation in this case, the transitions can only occur if 
the electron-photon interaction is accompanied by simultaneous electron 
scattering by phonons or static defects. Such indirect transitions are 
theoretically treated as second-order virtual processes via intermediate 
states. The intermediate states can be those
within the same quantum subband $e1$ as well as in other conduction or 
valence subbands. Figures~\ref{Figure06}(b) and (d) show possible absorption pathways
with intermediate states in the subband $e1$ and the excited subband $e2$.

The pathway, which is usually considered and determines the structure 
absorbance at normal incidence, involves intermediate states
within the subband $e1$. Such transitions (path\,I) are shown in 
Figs.~\ref{Figure06}(b) and (c) for the process, where the electron-photon 
interaction is followed by electron scattering [Fig.~\ref{Figure06}(b)]
and the inverted sequence process [Fig.~\ref{Figure06}(c)]. The matrix 
element of such kind of processes has the form
\begin{equation}\label{Me1}
M_{\mathbf{k}'\mathbf{k}}^{(1)} = \frac{eA}{c \omega m_{\perp}} \, \bm{e}\cdot(\bm{k}'-\bm{k}) V_{11} \:,
\end{equation}
where $\bm{k}$ and $\bm{k}'$ are the initial and final electron wave vectors, respectively, 
$e$ is the electron charge, $c$ the speed of light, $A$ the amplitude of 
the vector potential of the electromagnetic wave, which is related to the 
radiation intensity by $I=A^2 \omega^2 n_{\omega}/(2\pi c)$, 
and $V_{11}$ is the matrix element of electron scattering within the subband $e1$. 
We note that, while the matrix element in Eq.~(\ref{Me1}) is
odd in the wave vector, the absorption probability given by the squared matrix 
element is even in $(\bm{k}'-\bm{k})$. Thus, this type of processes alone does not introduce 
an asymmetry in the carrier distribution in $\bm{k}$-space and, consequently, does not yield an electric current.

In the geometry of oblique incidence, which is required for the CPGE in 
(001)-oriented structures, additional pathways with intermediate states 
in the excited subbands $e2$, $e3$ etc. 
also contribute to the radiation absorption.  Such virtual transitions 
(path II) via states in the $e2$ subband are sketched in Figs.~\ref{Figure06}(d) and (e). 
The matrix elements of the transitions with intermediate states 
in the subband $n$ ($n \neq 1$) have the form~\cite{Tarasenko07}
\begin{equation}\label{Me2}
M_{\mathbf{k}'\mathbf{k}}^{(n)} = i \frac{eA}{c \hbar} \left(
\frac{\varepsilon_{n1}}{\varepsilon_{n1}-\hbar\omega} -
\frac{\varepsilon_{n1}}{\varepsilon_{n1}+\hbar\omega} \right) e_z \, z_{n1} V_{1n} \:,
\end{equation}
where $\varepsilon_{n1}$
is the energy separation between the subbands at $\bm{k}=0$, $z_{n1} = \int\limits_{-\infty}^{+\infty} \phi_{n}(z) z \,
\phi_{1}(z)\: dz $ the
coordinate matrix element, $\phi_n(z)$ the function of size quantization, 
and $V_{1n}$ is the matrix element of intersubband scattering. 
Equation~(\ref{Me2}) shows that this type of indirect transitions 
is independent of $\bm{k}$ and, consequently, also does not result in an electric current.

The photocurrent emerges due to quantum interference of the virtual transitions 
considered above. Indeed, the total probability for the real optical transition $\bm{k} \rightarrow
\bm{k}'$ is given by the squared modulus of the sum of matrix elements describing individual pathways,
\[
W_{\bm{k}\prime\bm{k}} \propto |M^{(1)}_{\bm k^\prime \bm {k}} + \sum_{n \neq 1} M^{(n)}_{\bm k^\prime \bm {k}} |^2 = 
\]
\begin{equation}\label{W_kk}
|M^{(1)}_{\bm k^\prime \bm {k}}|^2 + | \sum_{n \neq 1}  M^{(n)}_{\bm k^\prime \bm {k}}|^2 +  2 \sum_{n \neq 1} \mathrm{Re}  [ M^{(1)}_{\bm k^\prime \bm {k}}
M_{\bm k^\prime \bm {k}}^{(n)*}] \,.
\end{equation}
Beside the probabilities of individual processes, it contains interference terms.
By using Eqs.\,(\ref{Me1}) and\,(\ref{Me2}) we derive for the last terms on the right-hand side of Eq.~(\ref{W_kk})
\[
\mathrm{Re}[ M^{(1)}_{\bm k^\prime \bm {k}} M_{\bm k^\prime \bm {k}}^{(n)*}] = i
\frac{e^2 A^2}{c^2 m_{\perp}} \frac{\varepsilon_{n1} z_{n1}}{\varepsilon_{n1}^2-(\hbar\omega)^2} V_{11} V_{1n}
\]
\begin{equation}\label{W_kk2}
\times [(k_x^\prime - k_x)  (e_z e_x^* - e_x e_z^*) + (k_y^\prime - k_y) (e_z e_y^* - e_y e_z^*)] \,.
\end{equation}
These terms are odd in the wave vector and, therefore, result
in different rates for the transitions to $\bm{k}'$ and $-\bm{k}'$. This
leads to an imbalance in the carrier distribution between $\bm{k}'$ and $-\bm{k}'$,
i.e. to an electric current $\bm{j}$. Such a difference in real
optical transition rates caused by constructive or destructive
interference of various pathways is illustrated in
Fig.\,\ref{Figure06}(a). Moreover, it follows from Eq.~(\ref{W_kk2}) that the 
sign of the interference terms is determined by components of the vector 
product $i [\bm{e} \times \bm{e}]$ and, therefore, by the radiation helicity due to
$i [\bm{e} \times \bm{e}] = \hat{\bm{e}} P_{circ}$.
Thus,  the imbalance of the carrier distribution in
$\bm{k}$-space and, consequently, the photocurrent reverse upon
switching  the light helicity.

To calculate the photocurrent we assume that electrons populate only the ground subband, they  
are elastically scattered by short-range defects,  and $\hbar\omega < \varepsilon_{21}$. 
Then, making an allowance for
transitions via all electron subbands, one can write for the
photocurrent~\cite{IvchenkoPikus_book,Tarasenko07}
\[
\bm{j} = e \frac{8\pi}{\hbar} \sum_{\bm{k},\bm{k}'}
[\tau_p(\varepsilon_{\bm{k}'}) \bm{v}_{\bm{k}'} -
\tau_p(\varepsilon_{\bm{k}}) \bm{v}_{\bm{k}} ]
[f(\varepsilon_{\bm{k}})-f(\varepsilon_{\bm{k}'})]
\]
\begin{equation}\label{j_gen}
\times \sum_{n \neq 1} 2 \mathrm{Re}[M^{(1)}_{\bm k^\prime \bm {k}} M_{\bm k^\prime \bm
{k}}^{(n)*}]
\delta(\varepsilon_{\bm{k}'}-\varepsilon_{\bm{k}}-\hbar\omega) \:,
\end{equation}
where $\tau_p$ is the momentum relaxation time, $\bm{v}_{\bm{k}} = \hbar
\bm{k}/m_{\perp}$ the electron velocity in the channel plane, 
$\varepsilon_{\bm{k}}=\hbar^2 \bm{k}^2/(2 m_{\perp})$ the electron 
kinetic energy measured from the subband
bottom, $f(\varepsilon_{\bm{k}})$ the function of equilibrium carrier 
distribution in the subband $e1$, and the factor 8
in Eq.~(\ref{j_gen}) accounts for the spin and valley degeneracy. For the 
electron scattering by short-range defects within the subband $e1$, the 
momentum relaxation time is given by $1/\tau_p = m_{\perp} \langle V_{11}^2 \rangle /\hbar^3$, 
where the angular brackets denote averaging over the impurity distribution. 

Finally, summing up Eq.~(\ref{j_gen}) over the wave vectors $\bm{k}$ and $\bm{k}'$ 
we derive the following expressions for components of the helicity dependent photocurrent:
\begin{equation}\label{j_final}
j_x /I = e \frac{4 \pi \alpha}{\omega n_{\omega}} \frac{\kappa
\hbar}{m_{\perp}} \sum_{n\neq 1} \frac{\langle V_{11} V_{1n} \rangle}{\langle V_{11}^2 \rangle}
\frac{\varepsilon_{n1}z_{n1}}{\varepsilon_{n1}^2-(\hbar\omega)^2}
N_s \,\hat{e}_y P_{circ} 
\end{equation}
and $j_y = - (\hat{e}_x/ \hat{e}_y) j_x$, 
where $\alpha=e^2/(\hbar c)$ is the fine-structure constant and $\kappa$ is 
the dimensional parameter, which is given by
\[
\kappa = \left. \int\limits_0^{\infty}
(1+2\varepsilon/\hbar\omega)
[f(\varepsilon)-f(\varepsilon+\hbar\omega)]\, d\varepsilon \right/
\int\limits_0^{\infty} f(\varepsilon) d\varepsilon \:
\]
and equal to 1 and 2 for the cases of $\hbar\omega \gg
\bar{\varepsilon}$ and $\hbar\omega \ll \bar{\varepsilon}$,
respectively, with the mean kinetic energy of carriers in the equilibrium $\bar{\varepsilon}$.

\subsection{The CPGE in MOSFETs on miscut surfaces}

In Si-MOSFETs fabricated on a miscut surface, the CPGE becomes possible even at normal 
incidence of radiation. This stems from the anisotropy of the electron dispersion in a valley 
together with the asymmetry of the confinement potential.~\cite{Magarill89,Olbrich09} 
The Hamiltonian describing electron states in the channel on a miscut surface in each 
of the low-energy valleys has the form (see, e.g. Ref.~[\onlinecite{IvchenkoPikus_book}])
\begin{equation}\label{Hamiltonian}
H = \frac{p_{x'}^2}{2m_{x'x'}} + \frac{p_{y'}^2}{2m_{y'y'}} +
\frac{p_{z'}^2}{2m_{z'z'}} + \frac{p_{y'} p_{z'}}{m_{y'z'}} + V(z') \:,
\end{equation}
where $\bm{p}$ is the momentum operator, $1/m_{\alpha\beta}$ is the tensor of the reciprocal effective
masses with the non-zero components $m_{x'x'}=m_{\perp}$, $1/m_{y'y'}=\cos^2\vartheta/m_{\perp}+\sin^2\vartheta/m_{\parallel}$,
$1/m_{z'z'}=\cos^2\vartheta/m_{\parallel}+\sin^2\vartheta/m_{\perp}$, 
$1/m_{y'z'}=(1/m_{\perp}-1/m_{\parallel})\cos\vartheta \sin\vartheta$, and $V(z')$ is the 
confining potential. The only nontrivial symmetry element, which does not change the 
Hamiltonian in Eq.~(\ref{Hamiltonian}) provided that $V(-z') \neq V(z')$, is the 
mirror reflection $x' \rightarrow -x'$. Thus, the Hamiltonian corresponds to the $C_s$ 
point-group symmetry that allows the photocurrent in the geometry of normal incidence of radiation. 

The eigen wave functions $\psi_{n\bm{k}}(\bm{\rho}',z')$ and energies $E_{n\bm{k}}$
of the Hamiltonian~(\ref{Hamiltonian}) are given by
\begin{equation}
\psi_{n\bm{k}}(\bm{\rho}',z') = \exp\left(i \bm{k}\cdot \bm{\rho}' -i  \frac{m_{z'z'}}{m_{y'z'}} \, k_{y'} z' \right) \phi_n(z') \:,
\end{equation}
\begin{equation}
E_{n\bm{k}} = E_{n} + \frac{\hbar^2
k_{x'}^2}{2 m_{x'x'}} + \frac{\hbar^2 k_{y'}^2}{2\tilde{m}_{y'y'}} \:,
\end{equation}
where $\bm{\rho}'=(x',y')$ is the in-plane
coordinate, $\phi_n(z')$ is the function of size quantization at $\bm{k}=0$,
$E_n$ is the energy of the bottom of the subband $n$, and $1/\tilde{m}_{y'y'}=1/m_{y'y'}-m_{z'z'}/m^2_{y'z'}$. 

Similarly to the CPGE in (001)-oriented structures, the microscopic origin of the 
CPGE is the quantum interference of virtual optical transitions with intermediate 
states in the ground and excited subbands. The latter processes are possible in 
structures on miscut surfaces even at normal incidence of radiation due to the 
presence of the off-diagonal component $p_{y'} p_{z'}/m_{y'z'}$ in the Hamiltonian, 
which couples the motion of carriers along the $y'$ and $z'$ axes. The matrix elements 
of virtual transitions via states in the ground and excited electron 
subbands have the form~\cite{Olbrich09}
\begin{equation}\label{M_vicinal}
M_{\bm{k}'\bm{k}}^{(1)} = \frac{eA}{c \omega} \left[
\frac{(k_{x'}'-k_{x'})\,e_{x'}}{m_{x'x'}} + \frac{(k_{y'}'-k_{y'})\,e_{y'}}{\tilde{m}_{y'y'}} \right] V_{11}
\:,
\end{equation}
\[
M_{\bm{k}'\bm{k}}^{(n)} = 2i \frac{eA}{c \hbar} \frac{m_{z'z'}}{m_{y'z'}}
\frac{\hbar \omega\, \varepsilon_{n1}z_{n1}'}{(\hbar\omega)^2 -
\varepsilon_{n1}^2} e_{y'} \,  V_{1n} \:.
\]

Finally, calculating Eq.~(\ref{j_gen}) with the matrix elements of Eq.~(\ref{M_vicinal}) 
and taking into account that 
$1/\tau_{p}=\sqrt{m_{x'x'}\tilde{m}_{y'y'}}\langle V_{11}^2 \rangle /\hbar^3$ 
in the case of short-range scattering, we derive
\[
j_{x'} /I = e \frac{4 \pi \alpha}{\omega n_{\omega}} \frac{\kappa
\hbar}{m_{y'z'}} \frac{m_{z'z'}}{m_{x'x'}} \sqrt{\frac{\tilde{m}_{y'y'}}{m_{x'x'}}}
\]
\begin{equation}\label{j_final_vicinal}
\times \sum_{n \neq 1}  \frac{\langle V_{11}V_{1n} \rangle}{\langle V_{11}^2 \rangle}
\frac{\varepsilon_{n1}z_{n1}'}{\varepsilon_{n1}^2-(\hbar\omega)^2} 
N_s \,\hat{e}_{z'} P_{circ} \:.
\end{equation}

According to the symmetry analysis, the helicity-dependent
photocurrent induced by normally-incident radiation arises to the
extent of the channel plane deviation from the plane $(001)$
together with the channel asymmetry. This follows also from
Eq.~(\ref{j_final_vicinal}), which demonstrates that the current $j_{y'}$
vanishes if the channel plane is parallel to $(001)$, where
$1/m_{y'z'}=0$, or for the symmetrical structure, where
$z_{n1}'\langle V_{11}V_{1n} \rangle =0$ for any $n$.

\section{Discussion} \label{sdiscussion}

Equations~(\ref{j_final}) and~(\ref{j_final_vicinal}) describe the main features of the circular photocurrent observed
in  electron channels on (001)-oriented and miscut silicon surfaces. In fact, it follows from Eq.~(\ref{j_final}) that
the CPGE in the (001)-oriented structure can occur only at oblique incidence and in the direction normal to the incidence plane. 
By contrast, in miscut structures, the circular photocurrent along $x^\prime$ can be excited even at normal
incidence [see Eq.~(\ref{j_final_vicinal})]. Such a behavior of the CPGE in respect to the light propagation 
direction and crystallographic orientation has indeed been observed in all samples under study.

The microscopic origin of the circular photocurrent reversal upon variation of  the 
energy separation between the subbands from $\varepsilon_{21}<\hbar\omega$ to $\varepsilon_{21}>\hbar\omega$ 
can be also clarified from Eqs.~(\ref{j_final}) and~(\ref{j_final_vicinal}).
In the vicinity of the intersubband resonance, the spectral dependence of the photocurrent
is given by $1 / [\varepsilon_{n1}^2 - (\hbar\omega)^2]$, which stems from the matrix elements describing virtual
transitions via the excited subbands, see Eqs.\,(\ref{Me2})~and \,(\ref{M_vicinal}). Thus, the photocurrent should increase drastically 
and undergo spectral inversion at $\hbar\omega \approx \varepsilon_{n1}$. In experiment, the resonant condition
is obtained by tuning the intersubband separation with the gate voltage while the photon energy $\hbar\omega$ 
is fixed. Since at low temperatures only the ground subband is populated, the photocurrent inversion upon variation
of $V_g$ is observed for $\hbar\omega \approx \varepsilon_{21}$, see Fig.~\ref{figure01}. 
While Eqs.~(\ref{j_final}) and~(\ref{j_final_vicinal})
yield sharp spectral resonances, in real structures the dependence smooths because
of broadening, but the inversion remains. 

The magnitude of the CPGE detected in the transistor on the miscut surface 
with $\vartheta = 9.7^\circ$ for $\hbar\omega = 8.4$\,meV and $V_{g}$\,=\,3\,V is $J_x/P \sim 1$\,nA/W, yielding the current
density $j_x/I \sim 0.1$\,nA\,cm/W. The same order of magnitude is obtained
from Eq.\,(\ref{j_final_vicinal}) for the structure with the miscut angle $\vartheta =
9.7^\circ$, the carrier density $N_s= 5 \times 10^{11}$\,cm$^{-2}$ ($V_{g}$= 3
V), the channel width $a = 8$\,nm 
and the structure asymmetry degree
$\langle V_{11}V_{12} \rangle /\langle V_{11}^2 \rangle =
10^{-2}$.

In deriving Eqs.~(\ref{j_final}) and~(\ref{j_final_vicinal}) we considered 
the intrasubband optical transitions within the ground subband assuming that
the momentum relaxation time $\tau_p$ is independent of the electron energy. 
Such an approximation is reasonable when the kinetic energy of photoexcited carriers
is smaller than $\varepsilon_{21}$.
For electrons generated with the energy $\varepsilon_{\bm{k}}
>\varepsilon_{21}$ the momentum relaxation time gets shorter
due to the additional relaxation channel caused by intersubband
scattering. Consequently,  one can expect that the magnitude of the  CPGE 
current is smaller at $\hbar\omega > \varepsilon_{21}$ (low $V_g$) than that at $\hbar\omega <
\varepsilon_{21}$ (high $V_g$). This can be responsible for the observed
asymmetry in the gate voltage dependence of the photocurrent in the vicinity of the intersubband resonance, see Fig.\,\ref{figure04}.
The gate voltage changes also the channel profile, which can be
taken into account assuming that the ratios $\langle V_{11}V_{1n} \rangle
/\langle V_{11}^2 \rangle$ and the coordinate matrix elements $z_{n1}$ depend on
$V_g$. At  $\hbar\omega \geq  \varepsilon_{21}$, possible contributions to the
CPGE due to intersubband optical transitions as well as 
scattering-induced broadening of the absorption peak should also be considered.\,\cite{Magarill89}

With increasing temperature the intersubband resonances are broadened 
and the excited subbands $e2$, $e3$ etc. become also occupied in the equilibrium. 
It leads to additional channels of the current formation including 
those with initial states in the excited subbands. As a result, both resonant LPGE and CPGE at 
$\hbar \omega \approx \varepsilon_{21}$ drastically decrease with temperature 
(see Fig.\,\ref{figure05}), and the spectral behavior of the photocurrent becomes more complicated.
We observe that at room temperature the CPGE current inversion with the gate 
voltage is detected only for the photon energy $\hbar \omega = 16.3$\,meV. 
Moreover, even in this case, the point of inversion does not correspond 
to $\hbar \omega = \varepsilon_{21}$.

\section{Summary}

In conclusion, we have studied the circular and linear photogalvanic effects in
Si-MOSFETs under excitation with linearly and elliptically polarized radiation in 
the terahertz frequency range. The behavior 
of photogalvanic effects upon variation 
of the radiation polarization state, wavelength, temperature and bias voltage have 
been investigated in transistors prepared on (001) precisely oriented as well as on miscut surfaces. 
The observed polarization properties of the photocurrent are well described by the phenomenological 
theory of the photogalvanic effects based on symmetry analysis of the studied structures.
Our experiments on Si-based structures reveal that the photon-helicity dependent
photocurrents can be generated in low-dimensional
semiconductors even with negligible spin-orbit interaction.  The mechanism of the photocurrent 
formation is based on the quantum interference of  different pathways contributing to the radiation absorption. 
The microscopic theory of the circular photogalvanic effect under intrasubband (Drude-like) 
absorption in the ground electron subband has been developed being in  a good agreement with the experimental findings.

\acknowledgments We thank E.L.\,Ivchenko,  V.V.\,Bel'kov, and J.\,Kamann for helpful discussions.
The financial support from the DFG, the Linkage Grant of IB of BMBF at DLR, and the RFBF 
is gratefully acknowledged.  S.A.T acknowledges also the support from 
the Russian President Grant for young scientists (MD-1717.2009.2).

\newpage

\pagebreak

\end{document}